\documentclass[a4paper, 10 pt, conference]{ieeeconf}

\IEEEoverridecommandlockouts

\usepackage{url}
\usepackage{graphicx}
\usepackage{soul}

\begin{document}

\title{On Using Blockchains for Safety-Critical Systems}

\author{Christian Berger$^{1}$, Birgit Penzenstadler$^{2}$, and Olaf Dr\"{o}gehorn$^{3}$
\thanks{$^{1}$Christian Berger is with the Department of Computer Science and Engineering, University of Gothenburg, Sweden
        {\tt\small christian.berger@gu.se}}%
\thanks{$^{2}$Birgit Penzenstadler is with the College of Engineering, California State University Long Beach, USA 
        {\tt\small birgit.penzenstadler@csulb.edu}}%
\thanks{$^{3}$Olaf Dr\"{o}gehorn is with Harz University of Applied Sciences
        {\tt\small odroegehorn@hs-harz.de}}%
}

\maketitle

\begin{abstract}
Innovation in the world of today is mainly driven by software. Companies need to continuously rejuvenate their product portfolios with new features to stay ahead of their competitors. For example, recent trends explore the application of blockchains to domains other than finance. This paper analyzes the state-of-the-art for safety-critical systems as found in modern vehicles like self-driving cars, smart energy systems, and home automation focusing on specific challenges where key ideas behind blockchains might be applicable. Next, potential benefits unlocked by applying such ideas are presented and discussed for the respective usage scenario. Finally, a research agenda is outlined to summarize remaining challenges for successfully applying blockchains to safety-critical cyber-physical systems.
\end{abstract}

\section{Introduction}
Today's industries in various domains are becoming more and more driven by software as innovator. They range from web applications powering our increasingly digitalized daily lives to deeply embedded systems driving complex and safety-critical cyber-physical systems (CPS) as in, for example, self-driving vehicles. Companies need to continuously rejuvenate their product portfolio for adopting new ideas to remain competitive. A recent idea that is permeating from its original application domain of financial use cases are blockchains, where researchers and companies try to apply key ideas behind them to other domains.

CPS are found in various domains like automotive, smart energy systems, and home automation. Many systems therein are strictly regulated as they provide safety-critical functionality to their users. Such domains are usually more conservative in adopting fundamentally new ideas as many hurdles need to be overcome to unlock an idea's full potential.

\begin{figure}[t]
	\centering
	\includegraphics[width=0.8\columnwidth]{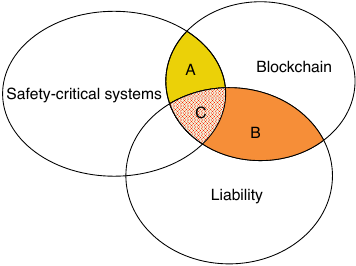}
	\caption{Key ideas behind blockchains might be applicable to safety-critical system as found in cyber-physical systems where liability is of importance. 
    }
	\label{fig:IntersectionSet}
\end{figure}

Figure \ref{fig:IntersectionSet} provides an overview of topic areas addressed in our research, which are currently potentially in conflict in industry: Today's companies that are developing products containing safety-critical components typically deal with the intersection set ``safety-critical systems'' and ``liability'' (for example, ISO-26262 in the automotive domain or insurance regulations in buildings and homes) to document what principles and state-of-practice were followed during the development and how the system was designed to reduce the risk of exposing its users to potentially harmful consequences. Such application scenarios, their benefits as well as their potential conflicts are analyzed and discussed in this paper. 

Our research is motivated by the aforementioned intersection areas, where key ideas behind blockchains might be applied to development and usage scenarios of safety-critical systems (intersection set ``A''). Furthermore, the design of blockchains might support certain areas for liability aspects (intersection set ``B''). Finally, key ideas behind blockchain are apparently of interest for the intersection of all three (labeled ``C''). Our goal with this paper is to explore the question: \textit{What research and engineering challenges discussed in literature could be addressed with key ideas behind blockchains?}

The main contributions of our work comprise the analysis of challenges for CPS in the three (a) domains automotive, (b) smart energy systems, and (c) home automation to identify benefits and challenges of adopting key ideas behind blockchains for deriving a research agenda focusing on CPS. While we limit our analysis to examples from the aforementioned areas, our research roadmap is not necessarily limited to them, but will contain topics that similarly affect other domains like aerospace. Furthermore, our study does not look into prototypical implementations of concrete blockchain technologies but discusses the benefits and advantages primarily from a conceptual perspective.

The rest of this article is structured as follows: Section \ref{sec:RelatedWork} presents and discusses related work. Section \ref{sec:Challenges} outlines three application domains and discusses specific challenges therein to motivate open research challenges to be tackled as part of a research agenda outlined in Section \ref{sec:ResearchAgenda}; the research agenda is provided in Section \ref{sec:Discussion} before the article is concluded in Section \ref{sec:Conclusion}.

\section{Related Work}
\label{sec:RelatedWork}

This section outlines the related work in blockchain technology, self-driving vehicles, smart energy systems, and home automation.

\subsection{Blockchain Technology}
Swan describes one of the uses of blockchain as ``Github for memory''~\cite[p.46]{7360255}. Blockchain technology is the secure decentralized computing ledger that enables Bitcoin and crypto-currencies. More profoundly, it is a next-generation global-scale decentralized infrastructure and mechanism for securely updating distributed computing nodes with ongoing consensus truth states~\cite[p.42]{7360255} -- recently also named as ``Internet-of-Values''. Blockchain 3.0 comprises applications beyond currency, finance, and markets -- particularly in the areas of government, health, science, literacy, culture, and art~\cite{swan2015blockchain}. 
Also the ICSE community has acknowledged the research potential to enable blockchain application development and Porru et al.~pointed out that the transactionality of blockchain requires major research efforts in software engineering~\cite{PorruPMT17} to fully understand and adopt its potential. 
Today's challenges for blockchains in general are their limited throughput, latency, size and bandwidth, security, inefficient resources usage, usability, and versioning, and hard forks resulting in multiple chains~\cite[p.82/83]{swan2015blockchain}. Furthermore, Caes et al.~\cite{caes2016robust} add scalability, compatibility, legal uncertainty, lack of industry standards, and cost efficiency.

\subsection{Self-Driving Vehicles}
The technology around self-driving vehicles has been demonstrated at large-scale for the first time during the 2004 \& 2005 DARPA Grand Challenges and with a focus on urban-like environments in the 2007 DARPA Urban Challenge (cf.~Rauskolb et al.~\cite{RBL+10}) ten years ago. Since then, technology start-ups and all major automotive original equipment manufacturers (OEMs) have started to make the technology more reliable and to prepare their product portfolio to provide autonomous driving for their customers\footnote{The California DMV currently lists more than 40 companies that have a permission to test their self-driving vehicles on public roads: \url{https://www.dmv.ca.gov/portal/dmv/detail/vehindustry/ol/auton\_veh\_tester}}. Recent examples from a US manufacturer of premium electric vehicles also demonstrate that newcomers in this domain started to adopt accelerated development and deployment at large scale even for safety-critical systems to better understand how the customers use their products in the field (cf.~Giaimo and Berger \cite{GB17}). Such over-the-air software updates for certified systems would require end-to-end documentation from a change request to its implementation and tests, up to its deployment and usage in the field.

However, following the ISO-26262 standard for vehicle functions from the area of autonomous driving bears many more challenges to overcome as mentioned by Martin et al.~\cite{MTBW17} and Spanfelner et al.~\cite{SRS+12}; for example, the rising area of machine-learning seems to be highly controversial when safety standards like ISO-26262 shall be followed. Salay et al.~\cite{SQC17} argue for instance that functionality, which is only partially specifiable like, for instance, algorithms for perceiving the surroundings, should be treated differently from fully specifiable ones. Similarly, V\"{o}st and Wagner \cite{VW17} argue that current literature about combining accelerated development (``continuous delivery'') with safety critical domains is scarce.
In our work, we explore opportunities and limitations from key principles behind blockchain to enable, for instance, accelerated development including continuous experimentation and thus, monitoring a product in the field and systematically modifying it after its initial deployment as the next step.

\subsection{Smart Energy Systems}
Current energy systems are transitioning towards smart grids that integrate an increasing list of energy sources, including wind, solar, water, gas, nuclear, and coal. \textit{Smart} means intelligent, neat, trim, stylish, or operating in automation; a grid is a network of electrical conductors that deliver electricity to certain points~\cite{tuballa2016review}. According to the Strategic Deployment Document for Europe's Electricity Networks of the Future, a Smart Grid is an electricity network that can intelligently integrate the actions of all users connected thereto; such actors include generators, consumers, and those that do both in order to efficiently deliver sustainable, economic, and secure electricity supplies~\cite{smartgrids2010smartgrids}.
Smart grids are an important component of a smart city~\cite{bulkeley} and potentially an application area for the sharing economy~\cite{kalathil}. The problem is that smart grids have not yet been proven in the context of the utility providers' desired specifications. ``Utilities are in critical need of a near-real-world environment, with real loads, distribution gear, and diverse consumption profiles, to develop, test, and validate their required smart grid solutions.''~\cite{farhangi2010path} Such an environment can be provided by a smart microgrid that would then be extended over time, to extensively test function bidding~\cite{kamyab2016demand} and to solve communication~\cite{tsampasis2016communication} as well as other challenges~\cite{tuballa2016review}.

\subsection{Home Automation} 
Another example for cyber-physical systems relates to Home Automation. The European Union identified that buildings consume 40\% of the energy and emit 36\% of CO2, being the largest end-user sector followed by transport (32\%), industry (24\%), and agriculture (2\%). The SMARTer 2030 report~\cite{smarter2030} by the Global eSustainability Initiative identified that the installation of building management systems (``smart home systems'') by occupants could offer a major opportunity to reduce the global CO2 emissions of buildings by a ratio of 15-30\%.
Although the possibilities arising from smart home technologies are promising, the market is highly fractionated and no set of standards for cross-vendor integration has emerged yet~\cite{icomp2016}. Therefore, we can see several open source projects emerging to offer the freedom to integrate different vendor technologies in a single smart home system. Despite the fact that the approach can lead to a fully integrated system, continuous integration of new vendor-specific technologies
are giving open source developer teams a hard time to keep the systems consistent and reliable. 
As insurers have already identified smart home technologies as a valid tool for preventing damages and insurance cases, they report the heterogeneity of available systems and the difficult integration of software versions as the major hampering reason for a solid business case~\cite{MITTechReview10-2016}.
Similar to the application of blockchain technology in smart homes for security and privacy~\cite{DBLP:journals/corr/DorriKJ16}, our research explores possibilities using blockchain technology to give house owners as well as insurers an on-demand overview of installed smart home components, their configuration, as well as their used software version. This would enable insurers, for example, to accept smart home systems when different vendors' products and their transactions would be transparently documented in an end-to-end manner.

\section{Challenges from Cyber-Physical Systems}
\label{sec:Challenges}

Cyber-physical systems denote systems where computation meets physical processes in various contexts ranging from classical control loops to, for example, monitor chemical processes, up to complex system-of-systems where several agents collaborate towards a joint goal, like, for instance harbor logistics. In the recent years, different research challenges have been suggested to be addressed from various perspectives.

Lee~\cite{Lee08} primarily highlights in his position paper limitations originating from the level of abstraction that is introduced with the computational view for CPS. In addition, he also points out that reliability and especially interfaces between CPS are critical to be designed and realized properly. In this regard, we see here a primary application domain for blockchain as transactions between different CPS or a CPS and its back-end in the cloud, for example, can be transparently documented in an end-to-end way.

Cheng et al.~\cite{CLG+09} present an extensive research roadmap primarily addressing self-adaptive systems. Several core issues that the authors identify for self-adaptive systems also concern CPS in general as both have to cope with uncertainty after deployment. The authors name, for instance, safe mechanisms to achieve adaptation at runtime. While they mainly look at self-adaptation, such type of modification at runtime can be found with continuous integration and continuous deployment as well as it is on purpose aiming at system modification. The authors suggest, for example, proof-carrying code (PCC) as an instrument to ensure that parts of a software unit adhere to certain safety rules. When using blockchains, for example, such safety rules could be stored thereon to document what types of rules were in effect at a certain point in time to cope with liability claims in case of system failures for example.

Furthermore, Cheng et al.~\cite{CLG+09} have identified challenges for self-adaptive systems, where we also see application areas for blockchains that hold for CPS in general. For example, the authors name traceability from requirements to implementation for documentation; here, we clearly see a benefit of using key ideas from the technology behind blockchain to capture all transactions that turn requirements into code to also address liability concerns again to cope with auditing.

Sanislav and Miclea~\cite{SM12} are concerned in their work with challenges and research areas for CPS in general. They look into the domains of aviation, automotive, energy, and healthcare. They have identified reliability and security as a prevailing topic, which is very important among all of them. Here, blockchain can serve as an instrument to address security when, for instance, interacting agents communicate with each other and need to log their interacting transactions.

The research agenda CPS by Giaimo and Berger \cite{GB15} also lists relevant challenges from CPS in general, where we foresee application areas for blockchains. The authors name, amongst others, safety and dependability of systems at runtime; as elaborated before, the auditing feature of blockchain could come into play here as well. Similarly, evolution of CPS can be supported by blockchain due to transparent documentation of transactions that model a system's evolution.

The possible application cases that we have described so far address, in general, documentation topics. Thereby, audits or certification of safety functionality can be tackled from an end-to-end perspective assuming an appropriate tool support. Today's rise of machine learning approaches, though, requires also a seamless documentation of the training and validation stages during the development, especially after a system's deployment to the field when more data and situations could be collected to modify a neural network (NN) retroactively: In cases where a system would expose unwanted side-effects or a system failure, the complete documentation of any transactions around a NN can address liability concerns.

\subsection{Selected Challenges from Self-Driving Vehicles}
\label{sec:selfdrivingvehicles}
Compared to the aforementioned challenges, the development of self-driving vehicular functionality is specifically influenced from the ISO-26262 standard concerning safety-critical functionality for road vehicles to, for instance, address liability concerns. Apparent application scenarios for blockchain would typically address end-to-end documentation of and about development artifacts. For example, changing one software unit on a self-driving vehicle could require a modification to the blockchain describing all software units present in the distributed system. Other software units could only accept messages sent from the updated software unit if these messages are signed with the most recent hash value corresponding to the head of the blockchain; otherwise, such messages could simply be ignored. Thereby, after-market modifications to a complex system-of-system could also be regulated.

The use of vehicle-to-infrastructure or vehicle-to-vehicle (V2X) communication could also benefit from using ideas from blockchain to improve the trustworthiness among communication parties: Using a V2X infrastructure could be combined with a blockchain that documents credibility of its communication participants, where several observers transparently document whether communicated intentions of a participating unit match with a participant's observed intentions; as soon as observed actions deviate from intended behavior of a participant, its credibility score would decrease in the blockchain-based documentation. The consequences of this unmotivated deviating behavior, for example, could be penalized from an Intelligent Transportation System (ITS) to reject any requests from accessing a shared resource like a fast-lane on highways or with an increased taxation.

\subsection{Selected Challenges from Smart Energy Systems}
There is a number of challenges in the smart energy systems domain, the two most important aspects being privacy~\cite{bekara2014security,rawat2015cyber} and security issues~\cite{crispim2014smart,xenias2015uk}.
The community needs to design strong data encryption schemes, and security solutions vary among domains~\cite{wang2013cyber}.

Becker et al.~\cite{becker2016requirements} advocate for the importance of requirements engineering systems that support the sustainability of human lifestyle on Earth. They point out that one of the challenges is to have the buy-in and trust from all involved stakeholders. A similar trust issue comes up when talking about emission trading in the form of smart contracts~\cite{alharby2017blockchain}. One potential solution could involve small-scale emission certificate trading using blockchain technology (similar to micro credits in Africa and India, cf.~\url{http://www.kiva.org}).

\subsection{Selected Challenges from Home Automation}
Smart Home systems are typically a system-of-systems and therefore consist of a multitude of different software artifacts. In the aforementioned case, where insurers are willing to support Smart Home installations for better home insurances, keeping track of the correct use and maintenance of the insured valuables (e.g., by smoke detectors, heating control systems, or similar), the used code in the different components, their configurations as well as their combination is part of liability concerns. In case of damage, the insurance company needs to be sure that smart home system itself has not caused or supported the damage, not been tampered with, that all measures previously agreed on have been taken to prevent the damage, and that the extent of the damage has been limited as good as possible. In order to ensure these aspects, several application scenarios for blockchains can be derived like keeping track of software-versions and configurations of the different artifacts. In the previous example an end-to-end documentation of the deployed software artifacts can be ensured by the blockchain technology as well as their individual configuration in the specific use case.

But beyond keeping track of installed and modified software versions, blockchains can also be used to monitor the transactions and messages exchanged between the different smart home components. This way, it can be tracked if users have been warned about critical states or have been informed to take specific actions in order to avoid severe damage. A challenge within this scenario is for sure the limited computational power of the different nodes within a smart home system but at least an integrating smart home server could perform more advanced blockchain operations. 

Furthermore Dorri, Kanhere, and Jurdak reported about the use of blockchain to tackle privacy and security issues in the Internet of Things, especially in the domain of interconnected Smart Homes \cite{DBLP:journals/corr/DorriKJ16}.
In order to foster the uptake of Smart Home Technology, blockchains can be used to keep track of the energy consumption of each home to reliably document fair use of resources by the inhabitants. Utilizing this, energy providers could either offer cheaper energy prices for more ``green'' families or a game-based competition between households could be started in order to motivate a more careful use of resources (cf.~\cite{gnauk2012leveraging,grossberg2015gamified}). This could ultimately lead to an approach of emission trading on a micro-scale within a street, a city, or a region.

\section{Research Agenda for Applying blockchain Ideas with Safety-Critical Systems}
\label{sec:ResearchAgenda}

\begin{figure*}
\centering
\includegraphics[width=0.8\textwidth]{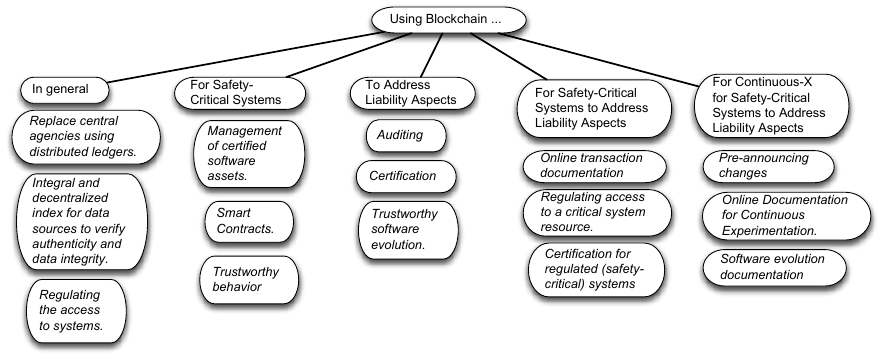}
\caption{Research agenda for development and usage scenarios for key ideas behind blockchain in Cyber-Physical Systems.}\label{fig:RA}
\end{figure*} 

Based on the challenges explored in the previous section, we outline several research topics originating from the overlapping circles in Figure~\ref{fig:IntersectionSet} describing potential application areas for key principles behind the blockchain technology. We suggest potential application scenarios; limitations of the current blockchain technology that might prevent a larger adoption, though, are further discussed in Section~\ref{sec:Discussion}.   

\ul{Using Blockchain -- General Case}

\textit{Replacing central agencies using the distributed and synchronized ledger idea.} One of the core ideas behind bitcoins is to make a regular bank account obsolete as the distributed blockchain behind bitcoins transparently contains all transactions between involved parties and a bank account balance is simply the summation of all transactions. The same principle can be used to replace existing centralized registrations like taxi license concessions or driver's licenses. 

\textit{Integral and decentralized index to existing data sources to verify authenticity and data integrity.} As the key idea behind blockchain is its immutability, this aspect can be exploited to establish and verify data integrity even for complex data sets. For example, sensor-readings over time from a test-drive of a self-driving vehicle could be stored by computing a hash-value per message that in turn is stored on the blockchain as the capacity on the blockchain is limited. A recipient of the complete dataset can then simply verify the integrity of the data recording by computing an arbitrarily chosen sequence of messages and verify their hash-values on the blockchain. Thereby, recent threats to pure hash-values applied to the entire dataset at large could be addressed\footnote{cf.~\url{https://goo.gl/lRCdTZ}}.

\textit{Regulating the access to systems.} The technology behind blockchain can also be used to replace current ``man-in-the-middle''-marketplaces like ride-sharing providers\footnote{cf.~\url{https://goo.gl/WfePp6}}. The blockchain would act as instrument to document supply and demand in a decentralized and distributed way to build the basis for service contracts for intelligent mobility. This would enable a cross-manufacturer marketplace to access intelligent mobility solutions in urban environments.

\ul{Using Blockchain for Safety-Critical Systems}

\textit{Management of certified software assets.} The development of a safety-critical system requires as of today strict documentation, for example, following the ISO-26262 in the automotive domain. Using blockchains, complete and full traceability of software assets that are used to create a software artifact including license management can be achieved to address auditing needs. 

\textit{Smart Contracts.} Recent blockchain-based technology like Ethereum embed a virtual machine that allows the execution of scripts. Such scripts could codify Smart Contracts as extension of the aforementioned topic, which can be verified at run-time (cf.~\cite{alharby2017blockchain}). Such Smart Contracts could be used, for example, for safety-critical software to verify certain properties of a blockchain-documented runtime environment before it is enabled to provide a contractually agreed level of service.

\textit{Trustworthy behavior.} Blockchains can be used to describe intended future behavior in a trustworthy way: An entity X can document its intended behavior on the blockchain and X's collaborators can verify if the actual observed behavior of X matches with the communicated behavior beforehand. Any deviation therefrom might result in a context-dependent penalty like exclusion of communication or rejected access to resources. This aspect, for example, can be used to enable a blockchain-based watchdog for safety-critical systems.

\ul{Using Blockchain to Address Liability Aspects}

\textit{Auditing.} Blockchains can be used to document requirements to implementation in a transparent, traceable, and non-falsifiable way. Thereby, the issues mentioned by V\"{o}st and Wagner~\cite{VW17} regarding how to handle artifacts around functional safety can be stored on a blockchain. In combination with the Smart Contract feature, automated auditing thereof is enabled to also pave the road towards true continuous integration even for safety-critical systems.

\textit{Certification.} Similarly, blockchains can be used to document a certified system structure by documenting all involved subsystems and software units. Thereby, unintended or uncertified system modifications are simply rejected and a system would simply refuse to get activated. Furthermore, due to the transparency in a distributed ledger, verification of a system with what has been certified before is enabled. Thereby, the remote removal of software not fulfilling a given certification after deployment would be transparent as hash-codes from individual software units would be document on an immutable blockchain and thus, any modification after the certification time point would be visible.

\textit{Trustworthy software evolution.} Today's software engineering is significantly powered by the version tracking software Git\footnote{cf. \url{https://www.openhub.net/repositories/compare}}. Compared to blockchain, both approaches implement Merkle trees. However, in contrast to blockchains, anyone can simply add a new hashed block to Git's Merkle tree. Blockchains when used for Bitcoins, for example, require a Proof-of-Work block to be added before any further transactions are considered verified and transparently documented across the distributed ledger. As such, a blockchain-based Merkle tree combined with Proof-of-Work blocks can help to realize auditable and transparent software evolution.

\ul{Using Blockchain for Safety-Critical Systems to Address Liability Aspects}

\textit{Online transaction documentation.} Blockchains can be used to document online transactions of a distributed system; exchanged messages (or a subset sufficient enough to recreate a certain system state) are stored on a blockchain. Transparently documenting (sub-)system interactions even, for example, for V2X scenarios can help to resolve liability questions in cases of system failure like a traffic accident.

\textit{Regulating access to a critical system resource.} As an extension of the aforementioned scenario, blockchains can be used to regulate and document the access to a critical system resource like V2X-supported traffic management at intersections. The consensus-principle behind blockchain could be exploited to prioritize who is scheduled to pass next through an intersection.

\textit{Certification for regulated (safety-critical) systems.} Publicly accessible and verifiable documentation of how safety-critical systems performed in standardized tests (e.g., vehicles' safety system in standardized New-Car-Assessment-Programs such as EuroNCAP) could be enabled through the blockchain technology.

\ul{Using Blockchains to Enable Continuous-X for Safety-Critical Systems to Address Liability Aspects}

\textit{Pre-announcing changes.} Similar to the aforementioned example of documented intended future behavior of interacting agents, blockchains can be used to prepare and document what is about to change as part of Continuous-X: Intended modifications to a system can be codified as Smart Contracts and verified before activation after deployment. Thereby, violating behavior can be spotted and documented in a transparent way withstanding auditing.  

\textit{Online Documentation for Continuous Experimentation.} Using blockchains, an experimental algorithm that is part of a Continuous Experiment (like, for example, an A/B test) and running in inert mode in the background is monitored by storing its incoming and outgoing data onto a blockchain to document its transaction history. Thereby, a system integrator can provide a safety-regulated instrument for its suppliers to change software running on a supplier's components after initial system deployment.

\textit{Software evolution documentation.} Blockchain can be used to document in a transparent way when suppliers or third-party entities change components in a system-of-systems after deployment. Thereby, certification violations and root-cause analysis can be facilitated in case of unwanted system behavior, malfunction, or system failure.

\section{Discussion}
\label{sec:Discussion}

The aforementioned research agenda focuses primarily on some core key principles behind the blockchain idea, namely documentation of transactions between interacting peers, distributed and decentralized storage, and immutability. These properties already enable critical application areas such as certification, smart contracting, or data integrity. However, cyber-physical systems might in many use cases require further properties that are either not or only partially addressed by blockchains as of today. In the following, we are discussing several challenges that we foresee when aiming for applying key principles behind blockchain to the domain of embedded and cyber-physical systems in particular.

The most critical limitation for embedded and cyber-physical systems in use cases that aim for online (or on-board) usage is the \textit{limited realtime capability}; the extension of the blockchain powering bitcoins is based on blocks that have been found using an increasingly hard computational task as Proof-of-Work. Embedded systems have only very limited power to act as Proof-of-Work providers. In addition, the computational effort requires a substantial amount of energy that might not be available on Internet-of-Things (IoT) devices that operate on battery for example.

Furthermore, due to the decentralized nature of blockchain, \textit{trust among nodes} is established using consensus about the longest valid blockchain. Therefore, blocks or the entire blockchain needs to be replicated among participants, which poses issues when either only limited storage capacity is available on a cyber-physical system, or when the network bandwidth is not sufficient or not present. The challenge, hence, is how a node could provide its service when a live network connection is a pre-condition therefor. 

Especially for complex systems such as self-driving vehicles as presented in Section \ref{sec:selfdrivingvehicles}, the \textit{amount of data to be processed} can easily reach several hundreds of megabytes per second. Thus, storing such data on a blockchain is challenging as, for example, the block size in use for bitcoins is limited to 1MB. While this could be modified to any arbitrarily chosen size, the replication feature as described before must be preserved. Therefore, small block sizes are rather favorable over large storage capacity. A potential approach, though, could be to simply store digests or hash values from a real data snippet instead of the raw data itself.

The \textit{block size limitation}, however, is also a limiting factor for the number of transactions that can be handled per block. Current estimations for Ethereum are, for example, around 20 transactions per second depending on the maximum available Ethereum's gas limit. This might seem enough for an application to realize a decentralized intersection management system to replace traffic lights, but this limit would be valid globally and hence, not useful in practice.

The hurdles to apply key ideas behind blockchains to cyber-physical systems as we foresee them originate mainly from the \textit{trade-off between decentralized design to achieve trust and consensus and number of participating nodes to benefit from the use of blockchains}. The more nodes need to replicate (parts of) a blockchain as well as use or contribute to a blockchain, the more important are solutions and strategies to increase realtime-capability and replication/storage efficiency.

\section{Conclusion}
\label{sec:Conclusion}

In this work, we analyzed challenges for safety-critical embedded and cyber-physical systems to address the research question, which of these challenges could be overcome when adopting key ideas behind blockchains. This technology's principles, such as transactionality, immutable evolution based on Merkle trees, and distributed and decentralized storage offer apparent applicability to the intersection set between safety-critical systems and liability concerns. We outlined various examples for possible application areas looking specifically at the borders within the intersection sets safety-critical systems, liability, and blockchains: For example, software evolution even initiated by sub-system suppliers could be made transparent and auditable using blockchains as documentation instrument; similarly, certification of safety-critical embedded systems could be made more easily accessible using blockchains to enable verification of a system's components; in combination with Smart Contracts, such system evolution and certification could be combined and even enforced at runtime. Furthermore, an accelerated development and deployment process (e.g., continuous integration and continuous deployment) for such systems might benefit from the transactionality of blockchains, as an accelerated development and release cycle requires a much more stringent way of documenting changes in an immutable and auditable way.

Despite potential application areas for key principles behind blockchains, current challenges that also present issues in the operation for Bitcoins need to be overcome. Especially for cyber-physical systems, real-time capability by finding better or complementary instruments for the proof-of-work idea is needed. Furthermore, we need a solution to address the storage limitations per block to avoid expensive and lengthy replication but yet ensure transparency and consensus.



\end{document}